# Comment on "Effective-range function methods for charged particle collisions"

Yu. V. Orlov*

*Skobeltsyn Institute of Nuclear Physics, Lomonosov Moscow State University, Russia*

**Abstract**

The authors of a recent paper [Phys. Rev. C 97(2018) 044003] (Ref. [1]), D. Gaspard and J.-M. Sparenberg, attempt to consider an alternative method for the asymptotic normalization coefficients (ANC) calculating which differs from so-called approximate Δ method proposed earlier [Phys. Rev. C 96(2017) 034601] (Ref. [2]) by O.L. Ramírez Suárez, and J. M. Sparenberg. The abstract in Ref. [1], in essence, declares that the approximation used in Ref. [2], where a $\Delta_l$ function is introduced to fit the Coulomb-nuclear phase shifts $\delta_l^{(cs)}$ at low energies, is not correct, since this $\Delta_l$ function is not analytical at zero energy. In my view, this statement is erroneous. I believe that the origin of this mistake in Ref. [1] is due to adopting the re-normalized scattering amplitude form designed for resonant states, which is not valid for a bound state. It is shown in the fundamental work [Nucl. Phys. B 60 (1973) 443] Ref.[3] by Hamilton *et al.* that an effective-range function (ERF) $K_l(k^2)$ is a meromorphic in the physical sheet except for the poles of bound states on the imaginary positive axis of the momentum complex plane. In my paper [Nucl. Phys A, 1010 (2021) 122174] Ref. [4], a strict form of the re-normalized scattering amplitude is derived, which can be used for the analytical continuation to a bound state pole. It is found in Ref. 4 that the $\Delta_l$ function has analytical properties similar to those of the $K_l(k^2)$ function. Thus, both functions have no singularities at zero energy.

\* orlov@srd.sinp.msu.ru

## 1. Introduction

The paper in Ref. [1] is entitled "Effective-range function methods for charged particle collisions". This means that the standard effective-range function (ERF) method should serve as a starting point where the Coulomb-nuclear phase shifts $\delta_l^{(cs)}$ are input data. The re-normalized partial scattering amplitude of the ERF method in terms of the effective-range function $K_l(k^2)$ is written as

$$\tilde{f}_l = k^{2l}/\left[K_l(k^2) - 2\xi D_l(k^2)\, h(\eta)\right], \qquad (1)$$

where

$$K_l(k^2) = 2\xi D_l(k^2)[\Delta_l(k^2) + h_r(k^2)]. \qquad (2)$$

Here and below $l$ is the orbital momentum, $\xi = Z_1 Z_2 \mu\alpha = 1/a_B$, $\eta = \xi/k$ is the Sommerfeld parameter, $k = \sqrt{2\mu E}$ is the relative momentum; $\mu$ and $E$ are the reduced mass and the center-of-mass energy of the colliding particles with charge numbers $Z_1$ and $Z_2$ respectively; $\alpha$ is the fine-structure constant; $a_B$ is the Bohr radius. In the ERF method, experimental $K_l(k^2)$ function data are fitted in the form of an effective-range expansion or of the Padé–approximant that is used for an analytical continuation to a bound state or a resonance pole. Since the binding energy $\varepsilon$ is known with good accuracy, the value of $K_l(-\kappa^2)$, $\kappa = \sqrt{2\mu\varepsilon}$ at the pole point is fixed, which reduces the problem of fitting the experimental function to a simpler interpolation. One needs to calculate $h(k)$ for $k=i\kappa$ or complex $k$ values. The unit system with $\hbar = c = 1$ is used. The following notations are used in (1) and (2):

$$D_l(k^2) = \prod_{n=1}^{l}(k^2 + \xi^2/n^2),\quad D_0(k^2) = 1; \qquad (3)$$

$$\Delta_l(k^2) = \pi \cot \delta_l^{(cs)} / [\exp(2\pi\eta) - 1] = C_0^2(\eta) \cot \delta_l^{(cs)}. \tag{4}$$

$$C_0^2(\eta) = \pi / [\exp(2\pi\eta) - 1]. \tag{5}$$

The Coulomb part of the ERF, $h_r(k^2) = \text{Re} h(\eta)$, is defined by the function

$$h(\eta) = \Psi(i\eta) + (2i\eta)^{-1} - \ln(i\eta), \tag{6}$$

where $\Psi(i\eta)$ is the digamma function.

In Ref. 4 the Coulomb-nuclear phase shift $\delta_l^{(cs)}$, $\cot \delta_l^{(cs)}$ and a finite limit of the nuclear part, $\Delta_l(k^2)$, of the effective-range function (ERF) are derived for an arbitrary orbital momentum $l$ when $E \to 0$. It is shown that $\cot \delta_l^{(cs)}$ has an essential singularity at zero energy, but $\Delta_l(k^2)$ does not. The explicit finite limit of $\Delta_l(0)$ is found to be

$$\Delta_l(k^2) \to \Delta_l(0) = -(l!)^2 / 2a_l \xi^{2l+1}, \tag{7}$$

$$\Delta_0(0) = -a_B / 2a_0. \tag{8}$$

As a result, there is no singularity in the $\Delta_l(k^2)$ function at zero energy since $\Delta_l(k^2)$ has a finite explicit limit which depends on $l$ and the scattering length $a_l$. For the Padé approximation, a combination of fitted constants appears instead of $a_l$. These results are well known in the literature and generally accepted for the S-wave scattering. The analytical property of $\Delta_l(k^2)$ as a regular function enables the analytical continuation of the re-normalized scattering amplitude from the physical energy region to a bound state pole.

The product $Z_1 Z_2$ limits the application of the ERF method. While $Z_1 Z_2$ increases, $\Delta_l(k^2)$ becomes a negligible quantity compared to the Coulomb term $h_r(k^2)$. Consequently, in [4, 5], a new $\Delta h$ method is proposed and developed using the re-normalized amplitude in the form

$$\tilde{f}_l = k^{2l} / \{2\xi D_l(k_r^2)[\Delta_l(k^2) + h_r(k^2) - h(\eta)]\}, \tag{9}$$

which is derived directly from Eqs. (1, 2). The main point of the $\Delta h$ method is to fit the $\Delta_l(k^2)$ function according to the data on experimental phase shifts and then to separately make the analytical continuation of $\Delta_l(k^2)$ and the Coulomb terms to a bound state pole. The function

$$\Delta h(k) = h_r(k^2) - h(\eta) \tag{10}$$

is introduced. To calculate $\Delta h(k)$, the function $h_r(k^2)$ is analytically continued to $E<0$. For the continuation of $h_r(E) = \text{Re} h(\eta)$ to the negative energy region, one should know the analytical properties of the $h_r(E)$ function. The form of the expansion of $h_r(k^2)$ into a series of powers of $k^2$ follows from the well-known asymptotic series (see (**6.3.18**), (**6.3.19**) in Ref. [6]) which converge at $\eta^2 \to \infty$, $(a_B k)^2 \to 0$, $|\arg \eta| < \pi$):

$$h_r(k^2) \to h_{as}(\eta) = \frac{1}{12\eta^2} + \frac{1}{120\eta^4} + \frac{1}{252\eta^6} + \cdots \tag{11}$$

The first term in Eq. (11), $1/(12\eta^2) = (a_B k)^2 / 12$, sets the same limit for the functions $h_r(k^2)$ and $h(\eta)$ when $E \to 0$ ($h(\eta)$ is real at $E<0$). Then one has a limit when $\kappa a_B \to 0$:

$$\Delta h(i\kappa) \sim O[(\kappa a_B)^4]. \tag{12}$$

By definition, this asymptotic expansion converges only at the center point, which corresponds to $k^2=0$. This means that $h_{as}(\eta)$ generally differs from $h_r(k^2)$ and $h(\eta)$. The fact that $h_r(0) = 0$ is very important. Due to the asymptotic expansion, the analytical properties of $K_l(k^2)$ around zero energy are determined by those of $\Delta_l(k^2)$, and vice versa.

In the physical region, $h(\eta)$ is a complex function (see Eq.(6)). However, when $E$ changes its sign, $k \rightarrow i\kappa$, the function $h(\eta)$ becomes real. This fact is used in the standard ERF method in the procedure of the scattering amplitude analytical continuation from the physical energy to the bound state pole. It seems that the reason for the incorrect statement about the singularity of the $\Delta_l(k^2)$ at zero energy is due to the use of the expression for the re-normalized scattering amplitude in the form

$$\tilde{f}_l = k^{2l} / [2\xi D_l(k^2) (\Delta_l(k^2) - i\pi / [\exp(2\pi\eta) - 1])]. \tag{13}$$

Eq. (13) is convenient for the amplitude analytical continuation to the point of a resonance complex energy situated in the non-physical sheet, but not to the pole for the real energy of the bound state. From Eq. (13) it is impossible to obtain the real condition for the pole of a bound state. Ref. [4] shows clearly that $\Delta_l(k^2)$ is a continuous function that does not have singularities around zero energy. This conclusion is based on the fact that the function $K_l(k^2)$ is a meromorphic on the physical sheet (see [3]) and that the function $h_r(k^2)$ has no singularities around $E=0$, since there is the asymptotic expansion (11). The equation

$$\Delta_l(k^2) - i\pi / [\exp(2\pi\eta) - 1]) = 0, \tag{14}$$

which follows from (13) for finding resonances, has no real roots. Therefore the form (13) cannot be applied to bound states.

**Declaration of competing interest**

The author declares that he has no known competing financial interests or personal relation-ships that could have appeared to influence the work reported in this paper.

**Acknowledgements**

This work was partially supported by the Russian Science Foundation (Grant No. 19-02-00014). The author is grateful to Helen Mary Jones for editing the English.